\documentclass[aps,twocolumn,prd,nofootinbib,showpacs]{revtex4}
\usepackage{amsmath}
\usepackage{amssymb}
\usepackage{graphicx}
\usepackage{hyperref}
\usepackage{color}

\newcommand{\boldmathsymbol}[1]{{\ensuremath{\boldsymbol{#1}}}}

\newcommand{\MeV}{\textrm{MeV}}
\newcommand{\GHz}{\textrm{GHz}}

\newcommand{\ud}{\mathrm{d}}
\newcommand{\uc}{\mathrm{c}}
\newcommand{\ub}{\mathrm{b}}

\newcommand{\uiac}{\mathrm{iac}}
\newcommand{\uinf}{\mathrm{inf}}
\newcommand{\urad}{\mathrm{rad}}
\newcommand{\uend}{\mathrm{end}}
\newcommand{\ureh}{\mathrm{reh}}
\newcommand{\ueff}{\mathrm{eff}}
\newcommand{\unuc}{\mathrm{nuc}}
\newcommand{\uKL}{\mathrm{KL}}
\newcommand{\uPl}{\mathrm{Pl}}
\newcommand{\uMC}{\mathrm{MC}}

\newcommand{\Rreh}{R_{\ureh}}
\newcommand{\Rrad}{R_{\urad}}
\newcommand{\wreh}{\overline{w}_{\ureh}}
\newcommand{\rhoreh}{\rho_{\ureh}}
\newcommand{\rhoend}{\rho_{\uend}}
\newcommand{\rhonuc}{\rho_{\unuc}}
\newcommand{\Nreh}{N_{\ureh}}
\newcommand{\Nend}{N_{\uend}}

\newcommand{\DKL}{D_{\uKL}}

\newcommand{\calLeff}{\mathcal{L}_\ueff}
\newcommand{\calM}{\mathcal{M}}
\newcommand{\calB}{\mathcal{B}}

\newcommand{\Mp}{M_{\uPl}}
\newcommand{\eps}[1]{\epsilon_{#1}}
\newcommand{\epsstar}[1]{\eps{#1*}}
\newcommand{\Pstar}{P_*}

\newcommand{\thetac}{\boldmathsymbol{\theta}_{\uiac}}
\newcommand{\thetai}{\boldmathsymbol{\theta}_{\uinf}}

\newcommand{\CAMB}{\texttt{CAMB}}
\newcommand{\COSMOMC}{\texttt{COSMOMC}}
\newcommand{\ASPIC}{\texttt{ASPIC}}
\newcommand{\MULTINEST}{\texttt{MultiNest}}

\newcommand{\OmegaB}{\Omega_{\ub}}
\newcommand{\OmegaC}{\Omega_{\uc}}
\newcommand{\thetaMC}{\theta_{\uMC}}

\newcommand{\ucal}{\mathrm{cal}}
\newcommand{\udust}{\mathrm{dust}}
\newcommand{\uCIB}{\mathrm{CIB}}
\newcommand{\utSZ}{\mathrm{tSZ}}
\newcommand{\uPS}{\mathrm{PS}}
\newcommand{\ukSZ}{\mathrm{kSZ}}

\newcommand{\ycal}{y_\ucal}
\newcommand{\BBdust}{A_{B,\udust}}
\newcommand{\BBbetadust}{\beta_{B,\udust}}
\newcommand{\acibC}{A^{\uCIB}_{217}}
\newcommand{\xitSZCIB}{\xi^{\utSZ, \uCIB}}
\newcommand{\aszB}{A^{\utSZ}_{143}}
\newcommand{\apsA}{A^{\uPS}_{100}}
\newcommand{\apsB}{A^{\uPS}_{143}}
\newcommand{\apsBC}{A^{\uPS}_{143\times217}}
\newcommand{\apsC}{A^{\uPS}_{217}}
\newcommand{\aksz}{A^{\ukSZ}}
\newcommand{\kgalA}{A^{\udust TT}_{100}}
\newcommand{\kgalB}{A^{\udust TT}_{143}}
\newcommand{\kgalBC}{A^{\udust TT}_{143\times217}}
\newcommand{\kgalC}{A^{\udust TT}_{217}}
\newcommand{\galfEEA}{A^{\udust EE}_{100}}
\newcommand{\galfEEAB}{A^{\udust EE}_{100\times143}}
\newcommand{\galfEEAC}{A^{\udust EE}_{100\times217}}
\newcommand{\galfEEB}{A^{\udust EE}_{143}}
\newcommand{\galfEEBC}{A^{\udust EE}_{143\times217}}
\newcommand{\galfEEC}{A^{\udust EE}_{217}}
\newcommand{\galfTEA}{A^{\udust TE}_{100}}
\newcommand{\galfTEAB}{A^{\udust TE}_{100\times143}}
\newcommand{\galfTEAC}{A^{ \udust TE}_{100\times217}}
\newcommand{\galfTEB}{A^{\udust TE}_{143}}
\newcommand{\galfTEBC}{A^{\udust TE}_{143\times217}}
\newcommand{\galfTEC}{A^{\udust TE}_{217}}
\newcommand{\calA}{c_{100}}
\newcommand{\calC}{c_{217}}

\newcommand{\sbi}{\mathrm{SBI}}
\newcommand{\li}{\mathrm{LI}}

\hypersetup{
    bookmarks=true,         
    unicode=false,          
    pdftoolbar=true,        
    pdfmenubar=true,        
    pdffitwindow=false,     
    pdfstartview={FitH},    
    pdftitle={My title},    
    pdfauthor={Author},     
    pdfsubject={Subject},   
    pdfcreator={Creator},   
    pdfproducer={Producer}, 
    pdfkeywords={keyword1} {key2} {key3}, 
    pdfnewwindow=true,      
    colorlinks=true,       
    linkcolor=red,          
    citecolor=cyan,        
    filecolor=magenta,      
    urlcolor=blue,           
    linktocpage=true
}

\begin{document}

\title{Information gain on reheating: The one bit milestone}

\author{J\'er\^ome Martin} \email{jmartin@iap.fr}
\affiliation{Institut d'Astrophysique de Paris, UMR 7095-CNRS,
Universit\'e Pierre et Marie Curie, 98 bis boulevard Arago, 75014
Paris, France}

\author{Christophe Ringeval} \email{christophe.ringeval@uclouvain.be}
\affiliation{Centre for Cosmology, Particle Physics and Phenomenology,
  Institute of Mathematics and Physics, Louvain University, 2 Chemin
  du Cyclotron, 1348 Louvain-la-Neuve, Belgium}

\author{Vincent Vennin} \email{vincent.vennin@port.ac.uk}
\affiliation{Institute of Cosmology \& Gravitation, University of
  Portsmouth, Dennis Sciama Building, Burnaby Road, Portsmouth, PO1
  3FX, United Kingdom}

\date{\today}

\begin{abstract}
  We show that the Planck 2015 and BICEP2/KECK measurements of the
  cosmic microwave background (CMB) anisotropies provide together an
  information gain of $0.82 \pm 0.13$ bits on the reheating history
  over all slow-roll single-field models of inflation. This
  corresponds to a $40\%$ improvement compared to the Planck 2013
  constraints on the reheating. Our method relies on an exhaustive CMB
  data analysis performed over nearly $200$ models of inflation to
  derive the Kullback-Leibler entropy between the prior and the fully
  marginalized posterior of the reheating parameter. This number is a
  weighted average by the Bayesian evidence of each model to explain
  the data thereby ensuring its fairness and robustness.
\end{abstract}

\maketitle

\section{Introduction}
\label{sec:intro}

The release of the Planck 2015 data and its awaited CMB polarization
measurements have provided an unprecedented understanding of the
Universe content from the time of last scattering to
today~\cite{Adam:2015rua}. In addition to having solved the puzzle
concerning the origin of the $B$-modes detection by the BICEP2
telescope~\cite{Ade:2014xna, Mortonson:2014bja, Martin:2014lra,
  Flauger:2014qra, Ade:2015tva, Array:2015xqh}, the Planck 2015 data
have increased the effective number of modes, i.e., of measured
$a_{\ell m}$, by $55\%$ compared to the Planck 2013
release~\cite{Ade:2013ktc}. As shown in Ref.~\cite{Ade:2015xua}, this
translates into a reduction of uncertainties by typically one-sigma on
essentially all the cosmological parameters and, in particular, on the
scalar spectral index and tensor-to-scalar ratio. Meanwhile, all tests
performed so far by the Planck Collaboration to search for any
deviations with respect to the predictions of slow-roll single-field
inflation, such as presence of isocurvature modes or features, remain
null~\cite{Ade:2015lrj}. Although any measurements of primordial
non-Gaussianities would provide an overwhelming amount of information
onto nonlinear physics during cosmic Inflation, their nondetection
does not mean that the microphysics of the early Universe remains
invisible.

As discussed in Refs.~\cite{Martin:2006rs, Martin:2010kz} (see also
subsequent works~\cite{Easther:2011yq, Dai:2014jja,
  Rehagen:2015zma, Drewes:2015coa}), the epoch of reheating, the era
during which the vacuum energy of the inflaton is converted into
radiation, impacts the predicted values of the scalar spectral index
and tensor-to-scalar ratio for all inflationary models. As a result,
reducing uncertainties on the measured values of these two parameters
provides some nontrivial information on how reheating proceeds, even
if inflation is as simple as a slow-roll single-field
scenario~\cite{Martin:2014rqa}. All kinematic effects from the
reheating modify the CMB predictions through one, and only one,
parameter, $\Rreh$ defined by~\cite{Martin:2010kz}
\begin{equation}
\ln \Rreh = \ln \Rrad + \dfrac{1}{4} \ln\left(\dfrac{\rhoend}{\Mp^4}\right),
\end{equation}
where $\Rrad$ is a peculiar combination of reheating quantities
\begin{equation}
\begin{aligned}
\ln \Rrad & = \dfrac{\Delta \Nreh}{4} \left(3\wreh- 1 \right ) 
\\ & 
=
\dfrac{1-3\wreh}{12(1+\wreh)} \ln \left(\dfrac{\rhoreh}{\rhoend}
\right).
\end{aligned}
\end{equation}
Here, $\Delta \Nreh = \Nreh - \Nend$ is the duration of reheating in
number of $e$-folds, $\wreh$ is the \emph{mean} (i.e., averaged over
number of $e$-folds) equation of state parameter of the Universe during
this epoch, $\rhoreh$ is the energy density at the end of reheating,
defined to be the beginning of the radiation era, and $\rhoend$ is the
energy density at the end of inflation.

Within a given model of inflation, predicting the scalar spectral
index and tensor-to-scalar ratio requires us to specify a value for the
so-called rescaled reheating parameter
$\Rreh$~\cite{Martin:2006rs}. More specifically, once $\Rreh$ is
given, one can uniquely determine the number of $e$-folds before the end
of inflation at which an observable mode $k_*$ crosses the Hubble
radius~\cite{Price:2015qqb}, a quantity which is required to get the
actual values of the observed slow-roll
parameters~\cite{Martin:2013uma, Jimenez:2013xwa}. As discussed in
Ref.~\cite{Martin:2014nya}, in the absence of any information on how
reheating proceeds, one can use a set of minimal
assumptions. Reheating should occur after inflation and before
big-bang nucleosynthesis such that $\rhonuc < \rhoreh <
\rhoend$. Moreover, from the energy positivity conditions in general
relativity, and the definition of reheating which is not inflation,
one has $-1/3 <\wreh < 1$. If one makes the conservative choice
$\rhonuc^{1/4}=10\,\MeV$, one obtains
\begin{equation}
 -46 < \ln \Rreh < 15 + \dfrac{1}{3} \ln\left(\dfrac{\rhoend}{\Mp^4} \right).
\label{eq:Rrehprior}
\end{equation}
These bounds define a flat prior probability distribution $\pi(\ln
\Rreh)$ on the parameter $\ln \Rreh$, i.e., a Jeffreys' prior on the
rescaled reheating parameter $\Rreh$. Let us notice that, within a
given model of inflation, $\rhoend$ is a theoretical output and not an
additional parameter. This however implies that the prior on the
rescaled reheating parameter has an upper bound which is
model dependent. Along these lines, performing a CMB data analysis
within one model of inflation allows one to infer, among others, the
marginalized posterior probability distribution $P(\ln\Rreh|D)$ for
the parameter $\ln \Rreh$ under the data set $D$. The data are
constraining the reheating epoch as soon as the posterior $P$ ``is peaked''
compared to the prior $\pi$. In Ref.~\cite{Martin:2014nya}, we have
followed this route using the Planck 2013 data for almost $200$ models
of inflation taken from the \emph{Encyclopaedia Inflationaris}
collection~\cite{Martin:2014vha}. Over all these models, the Planck
2013 data have been shown to give a reduction factor of $40\%$ in
the ratio of standard deviations of $\ln\Rreh$ between the prior $\pi$
and the inferred posterior $P$. 

Although such a result shows that the Planck data yield significant
constraints on reheating, the ratio of standard deviations is wasting
some amount of information. For instance, if the posterior $P(\ln
\Rreh|D)$ is multivalued, i.e., has more than one maximum, the
standard deviation within the posterior could remain as large as
within a flat prior. In such a situation, the data would be
disfavoring some intermediate ranges of values and, thus, would
contain some information not accounted for in the standard deviation.
This is why in the present work we prefer to use the Kullback-Leibler
divergence~\cite{kullback1951} between the prior distribution $\pi$
and the posterior $P$,
\begin{equation}
\DKL = \int P(\ln\Rreh|D) \ln\left[
\dfrac{P(\ln\Rreh|D)}{\pi(\ln\Rreh)}  \right] \ud \ln\Rreh,
\label{eq:DKL}
\end{equation}
which precisely is a measure of the amount of information
provided by the data $D$ about $\ln \Rreh$~\cite{Kunz:2006mc,
  Liddle:2007fy}. This quantity is also the discrepancy measure
between the posterior $P$ and the prior $\pi$ when the prior is viewed
as an approximation of the posterior. Because the Kullback-Leibler
divergence is invariant under any reparametrizations $x=f(\ln\Rreh)$
and uses a logarithmic score function as in the Shannon's entropy, it
is a well-behaved measure of information~\cite{bernardo:2008}.

\begin{figure}
\begin{center}
\includegraphics[width=0.49\textwidth]{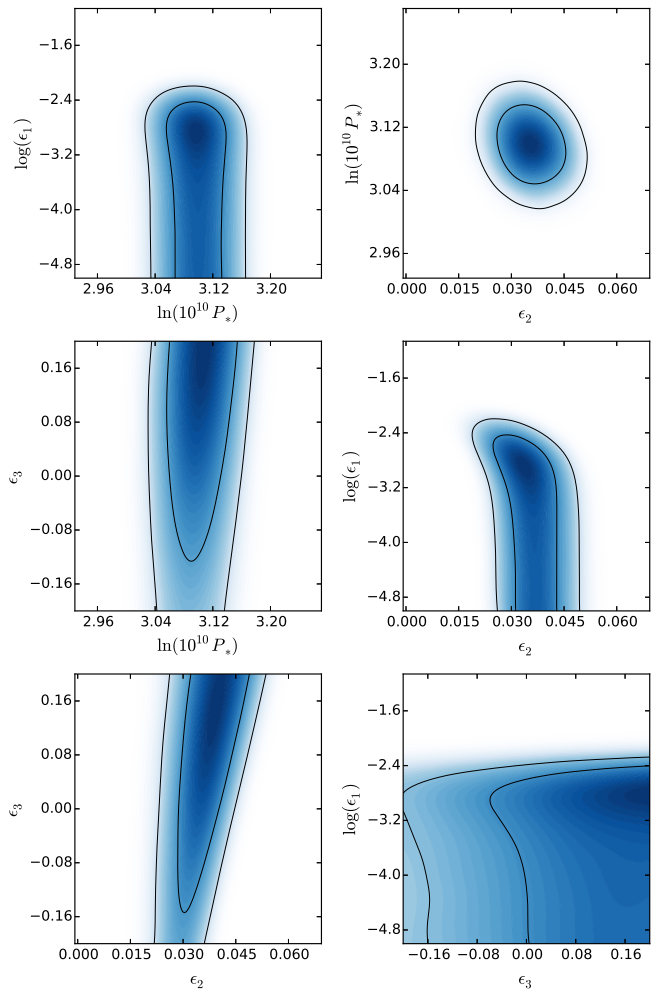}
\caption{One- and two-sigma confidence intervals associated wit the
  two-dimensional marginalized posterior distributions in the
  slow-roll parameter space
  $(\Pstar,\epsstar{1},\epsstar{2},\epsstar{3})$ from combined Planck 2015 and
  BICEP2/KECK data sets.}
\label{fig:sr2ndlog_bkppol}
\end{center}
\end{figure}

In the following, after having presented our data analysis method in
Sec.~\ref{sec:method}, we use Eq.~\eqref{eq:DKL} over the nearly
$200$ models of \emph{Encyclopaedia Inflationaris} to extract the
amount of information gained on the reheating parameter using the
Planck 2015 data~\cite{Aghanim:2015xee} complemented by the
BICEP2/KECK measurements of the B-modes
polarization~\cite{Array:2015xqh}. By using a base 2 logarithmic
function instead of the natural logarithm in Eq.~\eqref{eq:DKL}, the
information gain unit is the ``bit'' and this is our convention
in the rest of the paper. In Sec.~\ref{sec:results}, we show that
these two data sets combined give an information gain on reheating
equals to $\langle \DKL \rangle=0.83 \pm 0.13$. We have also performed
a new analysis of the Planck 2013 data, as in
Ref.~\cite{Martin:2014nya} but in terms of the Kullback-Leibler
entropy, to get $\langle {\DKL}_{13} \rangle = 0.55 \pm 0.14$. As a
result, Planck 2015 and BICEP2/KECK achieve an increase of $40\%$ more
information gain on reheating. The relevance of this number and its
implication for the Bayesian optimal design of future experiments are
discussed in the conclusion.

\begin{figure}
\begin{center}
\includegraphics[width=0.23\textwidth]{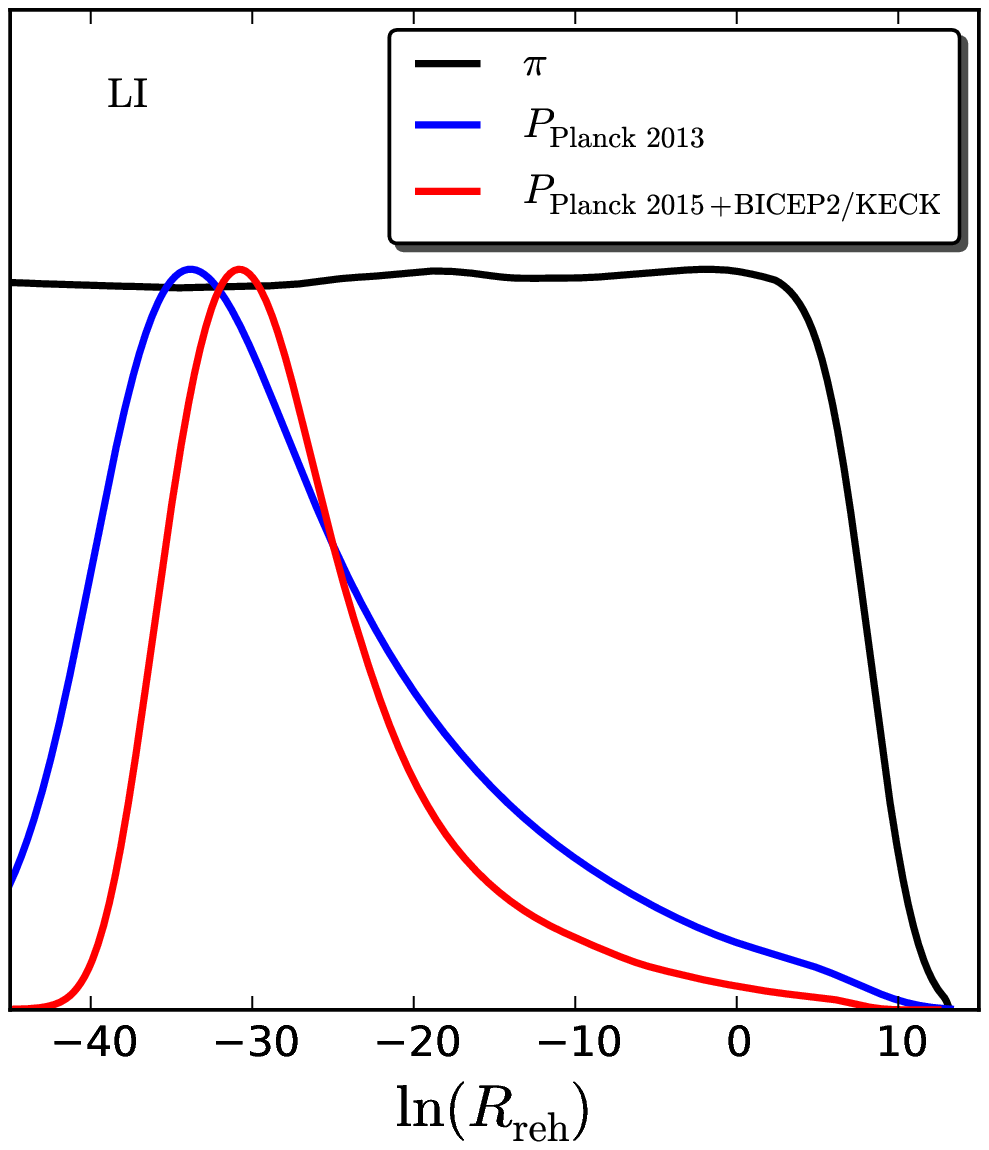}
\includegraphics[width=0.23\textwidth]{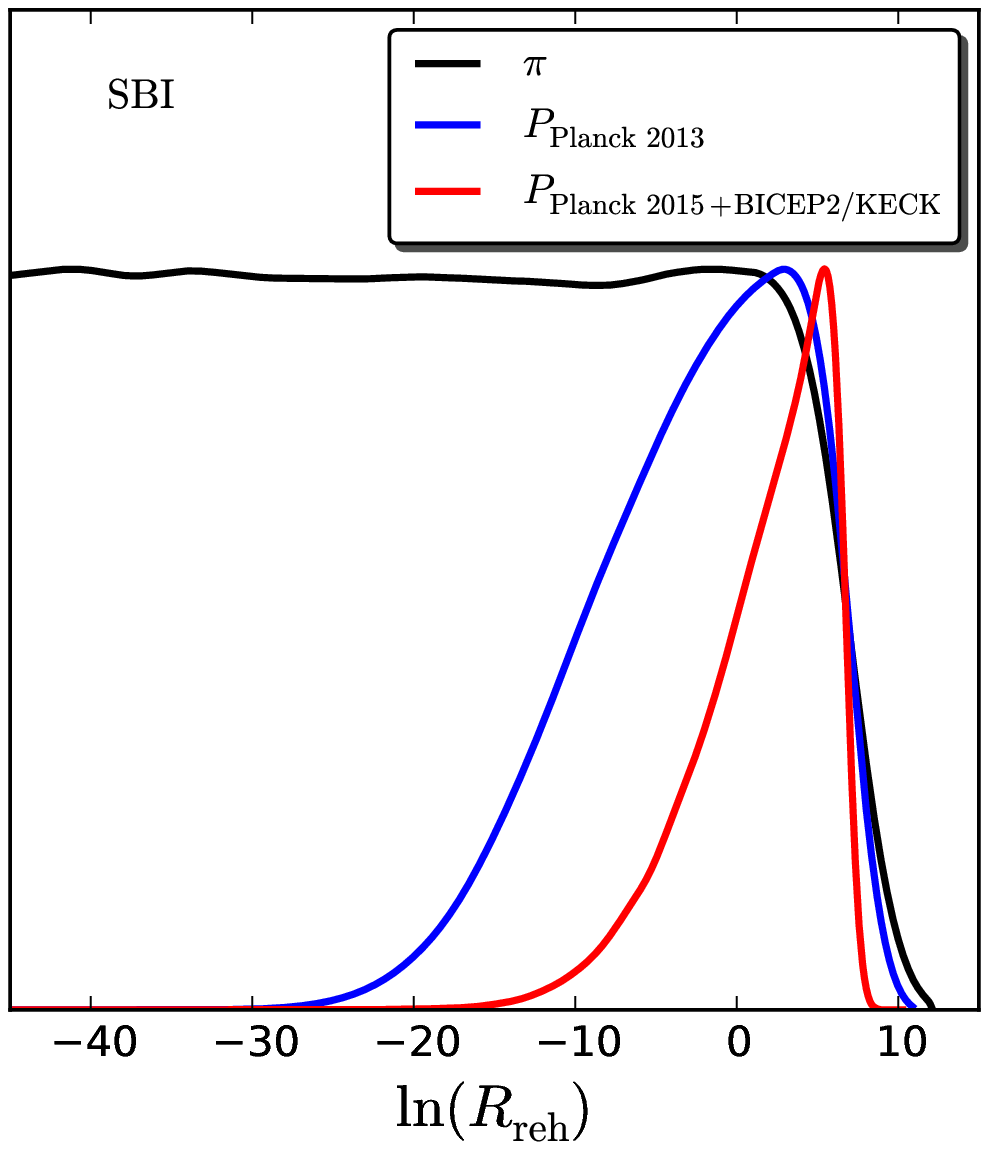}
\caption{Probability distributions (normalized to their maximum) for
  the rescaled reheating parameter $\Rreh$ associated with two of the
  $200$ models analyzed: loop inflation on the left ($\li$) and
  supergravity brane inflation on the right ($\sbi$), see
  Ref.~\cite{Martin:2014vha}. The black curve corresponds to the
  prior~(\ref{eq:Rrehprior}) and is not exactly flat since the upper
  bound of Eq.~(\ref{eq:Rrehprior}) is slightly model dependent. The
  marginalized posterior obtained from Planck 2013 data is displayed
  in blue and is to be compared to the more constraining posterior
  obtained from the Planck 2015 data with BICEP2/KECK (red curve).}
\label{fig:post}
\end{center}
\end{figure}

\begin{figure*}
\begin{center}
\includegraphics[width=0.49\textwidth]{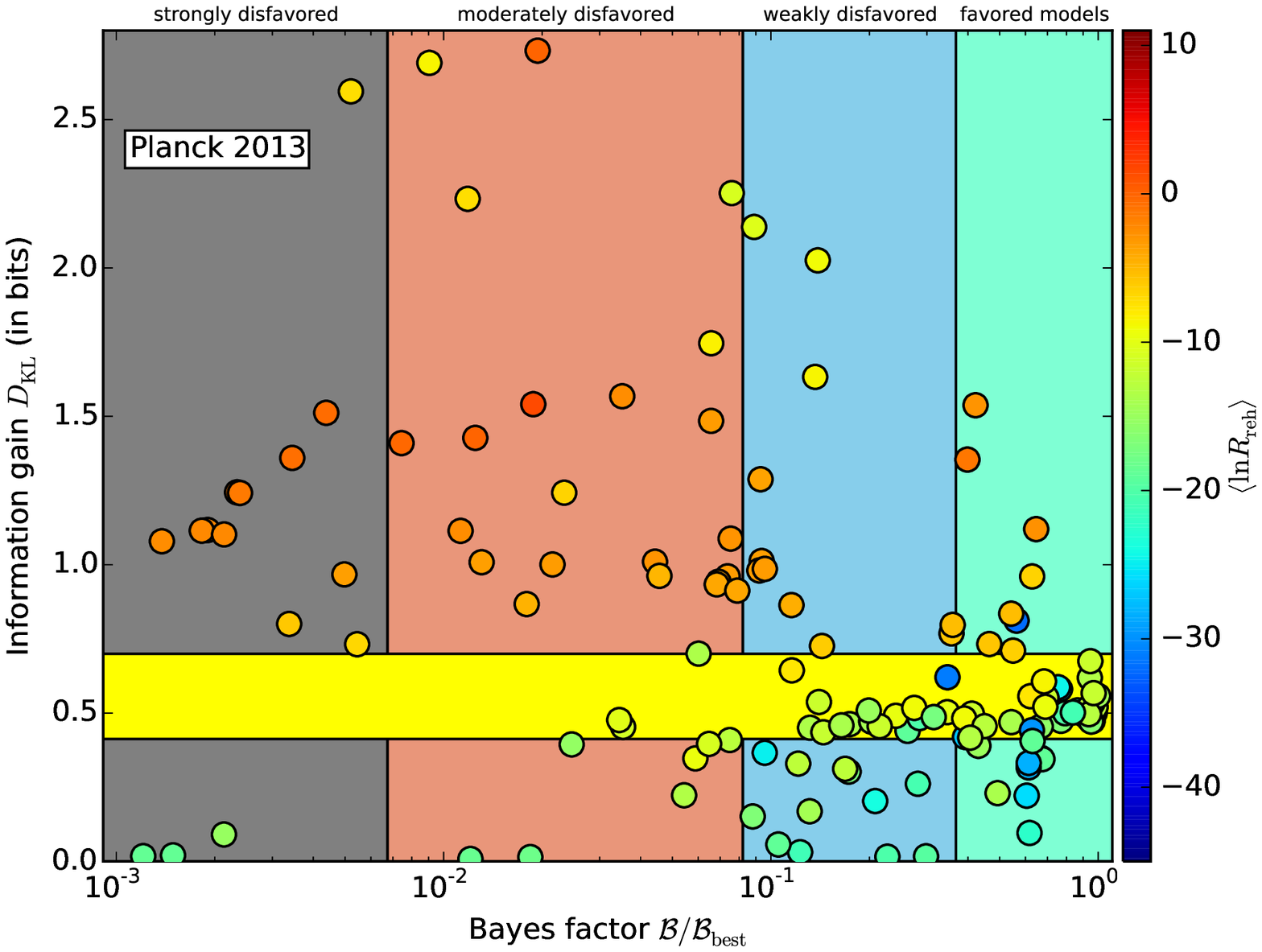}
\includegraphics[width=0.49\textwidth]{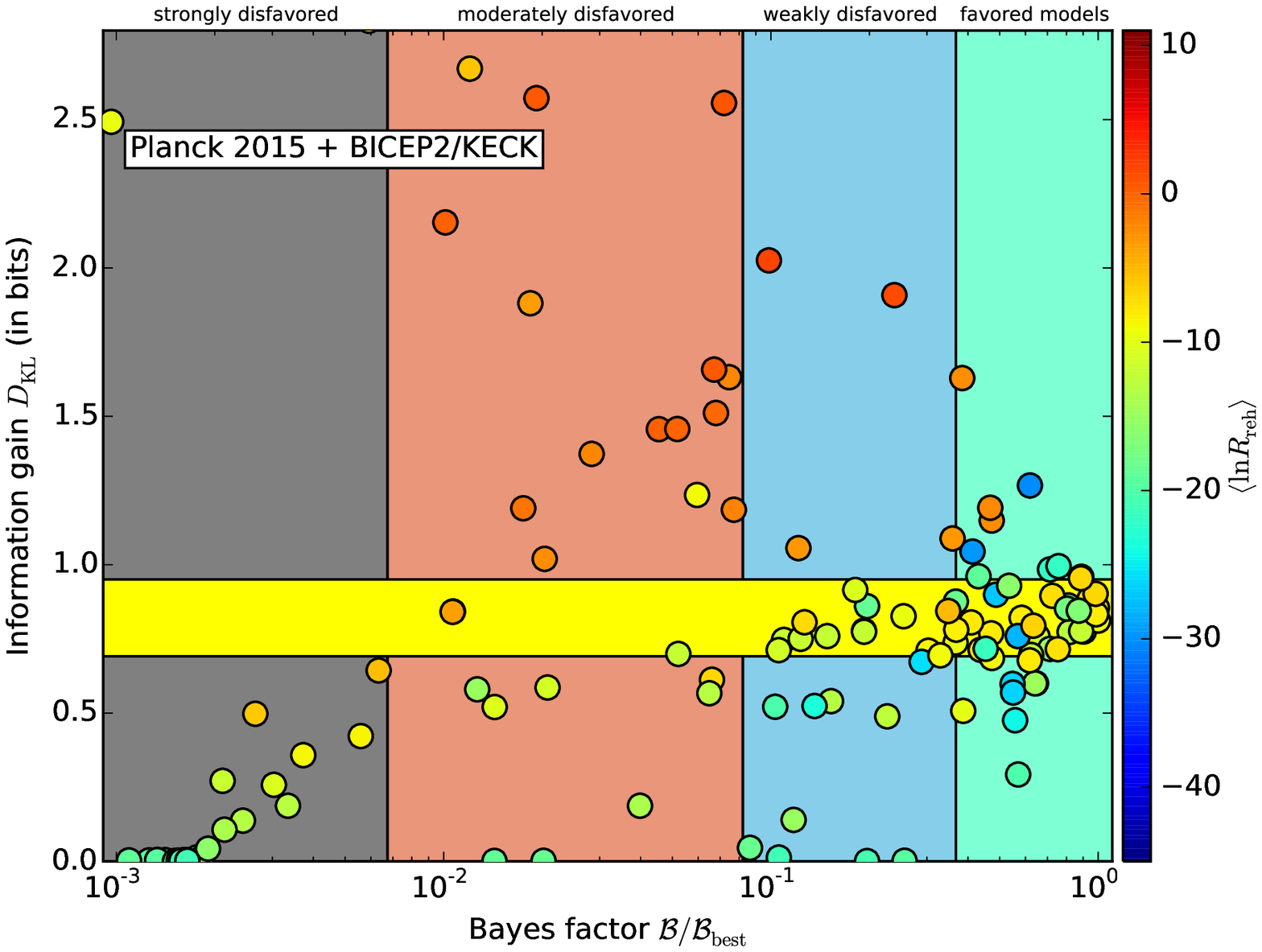}
\caption{Information gain $\DKL$ (in bits) given by Planck 2013 (left
  panel) and Planck 2015 with BICEP2/KECK (right panel) about the
  rescaled reheating parameter $\ln \Rreh$ as a function of the
  Bayesian evidence. Each circle represents one of the $200$ models of
  the \emph{Encyclopaedia Inflationaris} collection whose color traces
  the mean value of $\ln\Rreh$. The yellow band represents the
  one-sigma deviation around the mean value. For Planck 2015 and
  BICEP2/KECK, one gets $\langle \DKL \rangle = 0.82 \pm 0.13$. This
  corresponds to $40\%$ improvement compared to Planck 2013.}
\label{fig:onebit}
\end{center}
\end{figure*}

\section{Method}
\label{sec:method}

In order to perform the CMB data analysis of the hundreds of slow-roll
inflationary models at hand, we have followed the method described in
Ref.~\cite{Ringeval:2013lea} and applied in
Refs.~\cite{Martin:2013nzq, Martin:2014nya}. It consists first in the
determination of a machine learned effective likelihood $\calLeff$,
depending only on the slow-roll parameters $\lbrace\Pstar$,
$\epsstar{i}\rbrace$. The effective likelihood has been obtained by
marginalization of the joint Planck 2015 and BICEP2/KECK likelihood
over all the other parameters $\thetac$. These ones correspond to the
instrumental, astrophysical and cosmological parameters. One has
\begin{equation}
\calLeff(\Pstar,\epsstar{i}) \equiv \int
P(D|\thetac,\Pstar,\epsstar{i}) \pi(\thetac) \ud \thetac.
\label{eq:likeff}
\end{equation}
Within a given slow-roll model of inflation $\calM$, with theoretical
parameters $\thetai$, the quantities $\Pstar$ and $\epsstar{i}$ are
explicit (even if potentially complicated) functions of $\thetai$ and,
most importantly, of $\ln\Rreh$. As a result, from Bayes' theorem, the
posterior on $\ln \Rreh$ is given by~\cite{Ringeval:2013lea}
\begin{equation}
\begin{aligned}
& P(\ln \Rreh|D)  = \int P(\thetac,\thetai,\ln\Rreh|D) \, \ud \thetac
  \ud \thetai \\
& = \dfrac{\pi(\ln\Rreh)}{P(D|\calM)} \\
& \times 
\int \calLeff[\Pstar(\thetai,\ln\Rreh),\epsstar{i}(\thetai,\ln\Rreh)]
  \pi(\thetai) \ud \thetai,
\end{aligned}
\label{eq:post}
\end{equation}
where $P(D|\calM)$ is the global likelihood, which is proportional to
the Bayesian evidence $P(\calM|D)=P(D|\calM) \pi(\calM)$ of the model
$\calM$ to explain the data $D$.

In practice, we have used a modified version of the {\CAMB} and
{\COSMOMC} codes implementing the slow-roll primordial power spectra
at second order to perform a Markov-Chain-Monte-Carlo (MCMC) analysis
of the Planck 2015 and BICEP2/KECK data~\cite{Lewis:1999bs,
  Lewis:2002ah}. More precisely, we have used the public \texttt{plik}
likelihood provided by the Planck Collaboration, including the
polarization cross spectra $TE$ and $EE$ at large multipoles, together
with the BICEP2/KECK likelihood on $B$-modes based on the $217$
and $353\,\GHz$ maps~\cite{Ade:2015tva}. The parameter space
associated with $ \thetac$ contains the $4$ $\Lambda$CDM parameters
for a flat Universe: $\OmegaB h^2$, the density parameter of baryons
(times the squared reduced Hubble parameter $h$), $\OmegaC h^2$, of cold dark
matter, $\theta_{\uMC}$ related to the angular size of the sound
horizon at last scattering and $\tau$, the reionization optical
depth. There are $28$ additional parameters associated with
astrophysical signals and instrumental nuisances for both Planck 2015
and BICEP2/KECK. In total $\thetac$ belongs to a $32$-dimensional
parameter space and reads
\begin{equation}
\begin{aligned}
\thetac & = \left\{\OmegaB h^2, \OmegaC h^2,100 \thetaMC, \tau,
\right. \\ & \left. \ycal, \BBdust,\BBbetadust, \acibC, \xitSZCIB,
\aszB, \right. \\ & \left. \apsA, \apsB, \apsBC, \apsC,
\aksz, \kgalA, \kgalB, \right. \\ & \left. \kgalBC, \kgalC, \galfEEA,
\galfEEAB, \galfEEAC, \right. \\ & \left. \galfEEB, \galfEEBC, \galfEEC,
\galfTEA, \galfTEAB, \right. \\ & \left. \galfTEAC, \galfTEB, \galfTEBC,
\galfTEC, \calA, \calC \right\}.
\end{aligned}
\end{equation}
The parameters labeled by ``$A$'' refer to amplitude measurements
at various frequencies, $100$, $143$ and $217\,\GHz$ for
the temperature $T$ and polarization channels $E$, $B$, and all their
relevant cross-correlations. The astrophysical signals are associated
with unresolved point sources (PS), cosmic infrared background
(CIB), dust emission (dust) and kinetic (kSZ) and thermal
(tSZ) Sunyaev-Zeldowitch's effects. The parameter $\BBbetadust$ refers to the
spectral index of galactic dust emission in the $B$-mode polarization
channel and is required to correctly analyse the BICEP2/KECK data. The
other parameters encode calibration uncertainties. More details on
the meaning of these parameters can be found in
Ref.~\cite{Aghanim:2015xee}.

For our purpose, the prior distributions on the $\thetac$ parameters
have been chosen as specified in the Planck Collaboration's
paper~\cite{Ade:2015xua} while the priors on the slow-roll parameters
are as in Ref.~\cite{Martin:2013nzq}. Our MCMC exploration contains more
than $10^6$ samples which allows us to evaluate $\calLeff$ by
marginalization over $\thetac$. The machine learning algorithm used to
fit $\calLeff(\Pstar,\epsstar{i})$ is a modified quadratic Shepard's
method~\cite{Thacker:2010} as described in
Ref.~\cite{Ringeval:2013lea}. The two-dimensional marginalized
posteriors in the slow-roll parameter space $(\Pstar,\epsstar{1},
\epsstar{2}, \epsstar{3})$ are represented in
Fig.~\ref{fig:sr2ndlog_bkppol}.

For each models $\calM_i$ of the \emph{Encyclopaedia Inflationaris}
collection, we have obtained the posterior $P(\ln\Rreh|D)$ by using
the nested sampling algorithm {\MULTINEST}~\cite{Feroz:2007kg,
  Feroz:2008xx, Feroz:2013hea} on the effective likelihood $\calLeff$
to perform the marginalization of Eq.~\eqref{eq:post}. The slow-roll
functionals $\epsstar{i}(\thetai,\ln\Rreh)$ have been computed by
using the public library {\ASPIC}~\cite{Martin:2014vha} while the
priors for the $\thetai$ parameters have been set according to the
underlying theoretical setup as listed in Ref.~\cite{Martin:2013nzq}.

\section{Results}
\label{sec:results}

Let us now describe our main results. In Fig.~\ref{fig:post}, for
explanatory purposes, we have represented the posteriors of $\ln\Rreh$
for two (favored) models, named loop inflation ($\li$) and
supergravity brane inflation ($\sbi$), both from the Planck 2013 data
and from the Planck 2015 data with BICEP2/KECK. This figure
illustrates the gain of information between these two data sets as
well as the overall constraining power of CMB data on
reheating. Vanishing values of $\ln \Rreh$ correspond to a
radiation-like, or instantaneous, reheating scenario such that the
current data are actually ruling out such a scenario for $\li$ but
favoring it for $\sbi$~\cite{Martin:2014vha}. Of course, for other
models $\calM_i$, the posteriors on $\ln\Rreh$ are different and may
be peaked over large or small values, or not constrained at all. But
for all of them, $\DKL$ can be calculated.

In Fig.~\ref{fig:onebit}, we have represented by a circle each model
$\calM_i$ in the plane $(\calB,\DKL)$ where $\calB$ is the Bayes'
factor normalized to the best model. For a model $\calM_i$, assuming
noncommittal priors $\pi(\calM_i)=\pi(\calM_j)$, it is obtained from
the global likelihoods by
\begin{equation}
\calB_i \equiv \dfrac{P(\calM_i|D)}{\sup_{j}\left[P(\calM_j|D)\right]}
= \dfrac{P(D|\calM_i)}{\sup_{j}\left[P(D|\calM_j)\right]}\,.
\end{equation}
This figure shows that most of the models having large Bayes factors
are concentrated around values $\DKL \lesssim 1$ whereas disfavored
models may have $\DKL>2.5$. Such a correlation between information
gain and Bayes factors for disfavored models is not surprising. If a
model genuinely provides a bad fit to the data, the posterior of its
free parameters, including the reheating parameter, can be pushed to
the boundaries of their prior to be as good as it gets in improving
the fit. As a result, the model parameter space may end up being very
constrained while the model does not fit the data well compared to
others. Figure~\ref{fig:onebit} also shows some strongly
disfavored models with $\DKL=0$. These models are so far from the
favored region that even changing the reheating history does not help
to improve the fit to the data. For these reasons, a fair and robust
estimation of the information gain on reheating is given by the
average value of $\DKL$ in the space of all models
\begin{equation}
\langle \DKL \rangle = \sum_i P(\calM_i|D) \DKL(\calM_i)\simeq 0.82.
\label{eq:DKLplc2}
\end{equation}
As expected, it is weighted by the Bayesian evidence, namely the
probability of a model to explain the data: Disfavored
models weigh less than favored models. Similarly, we find for the
standard deviation
\begin{equation}
\sqrt{\langle \DKL^2 \rangle - (\langle \DKL \rangle)^2} \simeq 0.26.
\end{equation}

As mentioned in the introduction, we have performed the same analysis
for the Planck 2013 data and one gets $\langle {\DKL}_{13} \rangle =
0.55 \pm 0.14$. Therefore, Planck 2015 and BICEP2/KECK provide a
$40\%$ improvement in information gain compared to Planck
2013.

\section{Conclusion}
\label{sec:conclusion}

Because $\langle \DKL \rangle$ quoted in Eq.~\eqref{eq:DKLplc2} has
been derived over a significant number of models, it should be
representative of the whole information content of the current CMB
data about the reheating epoch within inflation. It is almost $1$
bit. Although this is a very modest number, $1$ bit is the amount of
information contained in answering ``yes'' or ``no'' to a given
question. As illustrated in Fig.~\ref{fig:post}, the question is about
the values of $\ln \Rreh$, and the current CMB data answer, on average,
whether $\ln\Rreh$ is large or small. According to
Shannon~\cite{Shannon:1951}, $1$ bit is also the typical amount of
information carried by one letter within the English language. As
such, one might argue that if the reheating scenario could be spelled,
the Planck 2015 and BICEP2/KECK data would allow us to know one
letter.

Within the deployment of the new generation of ground-based CMB
polarization telescopes, the soon to be operational Euclid
satellite~\cite{Scaramella:2015rra} and the much needed next
generation of CMB satellites~\cite{Matsumura:2013aja,core}, one can
only expect $\langle \DKL \rangle$ to dramatically increase in the
future. Measuring $\Rreh$ from the CMB would be a direct window onto
the microphysics after inflation, i.e., at energy scales which could
be as large as the grand unified theory energy scale. Moreover, it
would allow us to disambiguate two inflationary models having exactly
the same potential but not the same reheating history, as it is the
case for Starobinsky inflation and Higgs
inflation~\cite{GarciaBellido:2008ab, Terada:2014uia,
  Figueroa:2015rqa}, or for some curvaton
scenarios~\cite{Vennin:2015vfa,Vennin:2015egh}.

Another application of our result concerns the Bayesian optimal design
of future CMB measurements in which $\DKL$ could be used as a figure
or merit. In this situation, the best experimental setup is the one
maximizing the information gain $\DKL$. Because, for the reheating
parameter, $\DKL$ can be improved by increasing sensitivity in
\emph{all} the slow-roll parameters, this suggests that it is
certainly important to design future experiments to be as sensitive in
the scalar spectral index and scalar running as in the
tensor-to-scalar ratio. \\

\acknowledgments

V.V.'s work is supported by STFC Grants No. ST/K00090X/1 and
No. ST/L005573/1. C.R.'s work is supported by the Belgian Federal
Office for Science, Technical and Cultural Affairs.

\bibliography{biblio}

\begin{thebibliography}{44}
\expandafter\ifx\csname natexlab\endcsname\relax\def\natexlab#1{#1}\fi
\expandafter\ifx\csname bibnamefont\endcsname\relax
  \def\bibnamefont#1{#1}\fi
\expandafter\ifx\csname bibfnamefont\endcsname\relax
  \def\bibfnamefont#1{#1}\fi
\expandafter\ifx\csname citenamefont\endcsname\relax
  \def\citenamefont#1{#1}\fi
\expandafter\ifx\csname url\endcsname\relax
  \def\url#1{\texttt{#1}}\fi
\expandafter\ifx\csname urlprefix\endcsname\relax\def\urlprefix{URL }\fi
\providecommand{\bibinfo}[2]{#2}
\providecommand{\eprint}[2][]{\url{#2}}

\bibitem[{\citenamefont{Adam et~al.}(2015)}]{Adam:2015rua}
\bibinfo{author}{\bibfnamefont{R.}~\bibnamefont{Adam}} \bibnamefont{et~al.}
  (\bibinfo{collaboration}{Planck}) (\bibinfo{year}{2015}),
  \eprint{1502.01582}.

\bibitem[{\citenamefont{Ade et~al.}(2014{\natexlab{a}})}]{Ade:2014xna}
\bibinfo{author}{\bibfnamefont{P.}~\bibnamefont{Ade}} \bibnamefont{et~al.}
  (\bibinfo{collaboration}{BICEP2 Collaboration}),
  \bibinfo{journal}{Phys.Rev.Lett.} \textbf{\bibinfo{volume}{112}},
  \bibinfo{pages}{241101} (\bibinfo{year}{2014}{\natexlab{a}}),
  \eprint{1403.3985}.

\bibitem[{\citenamefont{Mortonson and Seljak}(2014)}]{Mortonson:2014bja}
\bibinfo{author}{\bibfnamefont{M.~J.} \bibnamefont{Mortonson}}
  \bibnamefont{and} \bibinfo{author}{\bibfnamefont{U.}~\bibnamefont{Seljak}},
  \bibinfo{journal}{JCAP} \textbf{\bibinfo{volume}{1410}}, \bibinfo{pages}{035}
  (\bibinfo{year}{2014}), \eprint{1405.5857}.

\bibitem[{\citenamefont{Martin et~al.}(2014{\natexlab{a}})\citenamefont{Martin,
  Ringeval, Trotta, and Vennin}}]{Martin:2014lra}
\bibinfo{author}{\bibfnamefont{J.}~\bibnamefont{Martin}},
  \bibinfo{author}{\bibfnamefont{C.}~\bibnamefont{Ringeval}},
  \bibinfo{author}{\bibfnamefont{R.}~\bibnamefont{Trotta}}, \bibnamefont{and}
  \bibinfo{author}{\bibfnamefont{V.}~\bibnamefont{Vennin}},
  \bibinfo{journal}{Phys.Rev.} \textbf{\bibinfo{volume}{D90}},
  \bibinfo{eid}{063501} (\bibinfo{year}{2014}{\natexlab{a}}),
  \eprint{1405.7272}.

\bibitem[{\citenamefont{Flauger et~al.}(2014)\citenamefont{Flauger, Hill, and
  Spergel}}]{Flauger:2014qra}
\bibinfo{author}{\bibfnamefont{R.}~\bibnamefont{Flauger}},
  \bibinfo{author}{\bibfnamefont{J.~C.} \bibnamefont{Hill}}, \bibnamefont{and}
  \bibinfo{author}{\bibfnamefont{D.~N.} \bibnamefont{Spergel}},
  \bibinfo{journal}{JCAP} \textbf{\bibinfo{volume}{1408}}, \bibinfo{pages}{039}
  (\bibinfo{year}{2014}), \eprint{1405.7351}.

\bibitem[{\citenamefont{Ade et~al.}(2015{\natexlab{a}})}]{Ade:2015tva}
\bibinfo{author}{\bibfnamefont{P.}~\bibnamefont{Ade}} \bibnamefont{et~al.}
  (\bibinfo{collaboration}{BICEP2, Planck}), \bibinfo{journal}{Phys. Rev.
  Lett.} \textbf{\bibinfo{volume}{114}}, \bibinfo{pages}{101301}
  (\bibinfo{year}{2015}{\natexlab{a}}), \eprint{1502.00612}.

\bibitem[{\citenamefont{Ade et~al.}(2016)}]{Array:2015xqh}
\bibinfo{author}{\bibfnamefont{P.~A.~R.} \bibnamefont{Ade}}
  \bibnamefont{et~al.} (\bibinfo{collaboration}{BICEP2, Keck Array}),
  \bibinfo{journal}{Phys. Rev. Lett.} \textbf{\bibinfo{volume}{116}},
  \bibinfo{pages}{031302} (\bibinfo{year}{2016}), \eprint{1510.09217}.

\bibitem[{\citenamefont{Ade et~al.}(2014{\natexlab{b}})}]{Ade:2013ktc}
\bibinfo{author}{\bibfnamefont{P.}~\bibnamefont{Ade}} \bibnamefont{et~al.}
  (\bibinfo{collaboration}{Planck Collaboration}),
  \bibinfo{journal}{Astron.Astrophys.} \textbf{\bibinfo{volume}{571}},
  \bibinfo{pages}{A1} (\bibinfo{year}{2014}{\natexlab{b}}), \eprint{1303.5062}.

\bibitem[{\citenamefont{Ade et~al.}(2015{\natexlab{b}})}]{Ade:2015xua}
\bibinfo{author}{\bibfnamefont{P.~A.~R.} \bibnamefont{Ade}}
  \bibnamefont{et~al.} (\bibinfo{collaboration}{Planck})
  (\bibinfo{year}{2015}{\natexlab{b}}), \eprint{1502.01589}.

\bibitem[{\citenamefont{Ade et~al.}(2015{\natexlab{c}})}]{Ade:2015lrj}
\bibinfo{author}{\bibfnamefont{P.~A.~R.} \bibnamefont{Ade}}
  \bibnamefont{et~al.} (\bibinfo{collaboration}{Planck})
  (\bibinfo{year}{2015}{\natexlab{c}}), \eprint{1502.02114}.

\bibitem[{\citenamefont{Martin and Ringeval}(2006)}]{Martin:2006rs}
\bibinfo{author}{\bibfnamefont{J.}~\bibnamefont{Martin}} \bibnamefont{and}
  \bibinfo{author}{\bibfnamefont{C.}~\bibnamefont{Ringeval}},
  \bibinfo{journal}{JCAP} \textbf{\bibinfo{volume}{0608}}, \bibinfo{pages}{009}
  (\bibinfo{year}{2006}), \eprint{astro-ph/0605367}.

\bibitem[{\citenamefont{Martin and Ringeval}(2010)}]{Martin:2010kz}
\bibinfo{author}{\bibfnamefont{J.}~\bibnamefont{Martin}} \bibnamefont{and}
  \bibinfo{author}{\bibfnamefont{C.}~\bibnamefont{Ringeval}},
  \bibinfo{journal}{Phys.Rev.} \textbf{\bibinfo{volume}{D82}},
  \bibinfo{pages}{023511} (\bibinfo{year}{2010}), \eprint{1004.5525}.

\bibitem[{\citenamefont{Easther and Peiris}(2012)}]{Easther:2011yq}
\bibinfo{author}{\bibfnamefont{R.}~\bibnamefont{Easther}} \bibnamefont{and}
  \bibinfo{author}{\bibfnamefont{H.~V.} \bibnamefont{Peiris}},
  \bibinfo{journal}{Phys.Rev.} \textbf{\bibinfo{volume}{D85}},
  \bibinfo{pages}{103533} (\bibinfo{year}{2012}), \eprint{1112.0326}.

\bibitem[{\citenamefont{Dai et~al.}(2014)\citenamefont{Dai, Kamionkowski, and
  Wang}}]{Dai:2014jja}
\bibinfo{author}{\bibfnamefont{L.}~\bibnamefont{Dai}},
  \bibinfo{author}{\bibfnamefont{M.}~\bibnamefont{Kamionkowski}},
  \bibnamefont{and} \bibinfo{author}{\bibfnamefont{J.}~\bibnamefont{Wang}},
  \bibinfo{journal}{Phys.Rev.Lett.} \textbf{\bibinfo{volume}{113}},
  \bibinfo{pages}{041302} (\bibinfo{year}{2014}), \eprint{1404.6704}.

\bibitem[{\citenamefont{Rehagen and Gelmini}(2015)}]{Rehagen:2015zma}
\bibinfo{author}{\bibfnamefont{T.}~\bibnamefont{Rehagen}} \bibnamefont{and}
  \bibinfo{author}{\bibfnamefont{G.~B.} \bibnamefont{Gelmini}},
  \bibinfo{journal}{JCAP} \textbf{\bibinfo{volume}{1506}}, \bibinfo{pages}{039}
  (\bibinfo{year}{2015}), \eprint{1504.03768}.

\bibitem[{\citenamefont{Drewes}(2016)}]{Drewes:2015coa}
\bibinfo{author}{\bibfnamefont{M.}~\bibnamefont{Drewes}},
  \bibinfo{journal}{JCAP} \textbf{\bibinfo{volume}{1603}}, \bibinfo{pages}{013}
  (\bibinfo{year}{2016}), \eprint{1511.03280}.

\bibitem[{\citenamefont{Martin et~al.}(2014{\natexlab{b}})\citenamefont{Martin,
  Ringeval, and Vennin}}]{Martin:2014rqa}
\bibinfo{author}{\bibfnamefont{J.}~\bibnamefont{Martin}},
  \bibinfo{author}{\bibfnamefont{C.}~\bibnamefont{Ringeval}}, \bibnamefont{and}
  \bibinfo{author}{\bibfnamefont{V.}~\bibnamefont{Vennin}},
  \bibinfo{journal}{JCAP} \textbf{\bibinfo{volume}{1410}}, \bibinfo{pages}{038}
  (\bibinfo{year}{2014}{\natexlab{b}}), \eprint{1407.4034}.

\bibitem[{\citenamefont{Price et~al.}(2016)\citenamefont{Price, Peiris, Frazer,
  and Easther}}]{Price:2015qqb}
\bibinfo{author}{\bibfnamefont{L.~C.} \bibnamefont{Price}},
  \bibinfo{author}{\bibfnamefont{H.~V.} \bibnamefont{Peiris}},
  \bibinfo{author}{\bibfnamefont{J.}~\bibnamefont{Frazer}}, \bibnamefont{and}
  \bibinfo{author}{\bibfnamefont{R.}~\bibnamefont{Easther}},
  \bibinfo{journal}{JCAP} \textbf{\bibinfo{volume}{1602}}, \bibinfo{pages}{049}
  (\bibinfo{year}{2016}), \eprint{1511.00029}.

\bibitem[{\citenamefont{Martin et~al.}(2013)\citenamefont{Martin, Ringeval, and
  Vennin}}]{Martin:2013uma}
\bibinfo{author}{\bibfnamefont{J.}~\bibnamefont{Martin}},
  \bibinfo{author}{\bibfnamefont{C.}~\bibnamefont{Ringeval}}, \bibnamefont{and}
  \bibinfo{author}{\bibfnamefont{V.}~\bibnamefont{Vennin}},
  \bibinfo{journal}{JCAP} \textbf{\bibinfo{volume}{1306}}, \bibinfo{pages}{021}
  (\bibinfo{year}{2013}), \eprint{1303.2120}.

\bibitem[{\citenamefont{Beltran~Jimenez
  et~al.}(2013)\citenamefont{Beltran~Jimenez, Musso, and
  Ringeval}}]{Jimenez:2013xwa}
\bibinfo{author}{\bibfnamefont{J.}~\bibnamefont{Beltran~Jimenez}},
  \bibinfo{author}{\bibfnamefont{M.}~\bibnamefont{Musso}}, \bibnamefont{and}
  \bibinfo{author}{\bibfnamefont{C.}~\bibnamefont{Ringeval}},
  \bibinfo{journal}{Phys.Rev.} \textbf{\bibinfo{volume}{D88}},
  \bibinfo{pages}{043524} (\bibinfo{year}{2013}), \eprint{1303.2788}.

\bibitem[{\citenamefont{Martin et~al.}(2015)\citenamefont{Martin, Ringeval, and
  Vennin}}]{Martin:2014nya}
\bibinfo{author}{\bibfnamefont{J.}~\bibnamefont{Martin}},
  \bibinfo{author}{\bibfnamefont{C.}~\bibnamefont{Ringeval}}, \bibnamefont{and}
  \bibinfo{author}{\bibfnamefont{V.}~\bibnamefont{Vennin}},
  \bibinfo{journal}{Phys. Rev. Lett.} \textbf{\bibinfo{volume}{114}},
  \bibinfo{pages}{081303} (\bibinfo{year}{2015}), \eprint{1410.7958}.

\bibitem[{\citenamefont{Martin et~al.}(2014{\natexlab{c}})\citenamefont{Martin,
  Ringeval, and Vennin}}]{Martin:2014vha}
\bibinfo{author}{\bibfnamefont{J.}~\bibnamefont{Martin}},
  \bibinfo{author}{\bibfnamefont{C.}~\bibnamefont{Ringeval}}, \bibnamefont{and}
  \bibinfo{author}{\bibfnamefont{V.}~\bibnamefont{Vennin}},
  \bibinfo{journal}{Phys. Dark Univ.} \textbf{\bibinfo{volume}{5-6}},
  \bibinfo{pages}{75} (\bibinfo{year}{2014}{\natexlab{c}}), \eprint{1303.3787}.

\bibitem[{\citenamefont{Kullback and Leibler}(1951)}]{kullback1951}
\bibinfo{author}{\bibfnamefont{S.}~\bibnamefont{Kullback}} \bibnamefont{and}
  \bibinfo{author}{\bibfnamefont{R.~A.} \bibnamefont{Leibler}},
  \bibinfo{journal}{Ann. Math. Statist.} \textbf{\bibinfo{volume}{22}},
  \bibinfo{pages}{79} (\bibinfo{year}{1951}),
  \urlprefix\url{http://dx.doi.org/10.1214/aoms/1177729694}.

\bibitem[{\citenamefont{Kunz et~al.}(2006)\citenamefont{Kunz, Trotta, and
  Parkinson}}]{Kunz:2006mc}
\bibinfo{author}{\bibfnamefont{M.}~\bibnamefont{Kunz}},
  \bibinfo{author}{\bibfnamefont{R.}~\bibnamefont{Trotta}}, \bibnamefont{and}
  \bibinfo{author}{\bibfnamefont{D.}~\bibnamefont{Parkinson}},
  \bibinfo{journal}{Phys. Rev.} \textbf{\bibinfo{volume}{D74}},
  \bibinfo{pages}{023503} (\bibinfo{year}{2006}), \eprint{astro-ph/0602378}.

\bibitem[{\citenamefont{Liddle}(2007)}]{Liddle:2007fy}
\bibinfo{author}{\bibfnamefont{A.~R.} \bibnamefont{Liddle}},
  \bibinfo{journal}{Mon. Not. Roy. Astron. Soc.}
  \textbf{\bibinfo{volume}{377}}, \bibinfo{pages}{L74} (\bibinfo{year}{2007}),
  \eprint{astro-ph/0701113}.

\bibitem[{\citenamefont{Bernardo and Smith}(2008)}]{bernardo:2008}
\bibinfo{author}{\bibfnamefont{J.~M.} \bibnamefont{Bernardo}} \bibnamefont{and}
  \bibinfo{author}{\bibfnamefont{A.~F.~M.} \bibnamefont{Smith}},
  \emph{\bibinfo{title}{Bayesian Theory}} (\bibinfo{publisher}{John Wiley \&
  Sons, Inc.}, \bibinfo{year}{2008}), pp. \bibinfo{pages}{105--164}.

\bibitem[{\citenamefont{Aghanim et~al.}(2015)}]{Aghanim:2015xee}
\bibinfo{author}{\bibfnamefont{N.}~\bibnamefont{Aghanim}} \bibnamefont{et~al.}
  (\bibinfo{collaboration}{Planck}), \bibinfo{journal}{Submitted to: Astron.
  Astrophys.}  (\bibinfo{year}{2015}), \eprint{1507.02704}.

\bibitem[{\citenamefont{{Ringeval}}(2014)}]{Ringeval:2013lea}
\bibinfo{author}{\bibfnamefont{C.}~\bibnamefont{{Ringeval}}},
  \bibinfo{journal}{Mon. Not. Roy. Astron. Soc.}
  \textbf{\bibinfo{volume}{439}}, \bibinfo{pages}{3253} (\bibinfo{year}{2014}),
  \eprint{1312.2347}.

\bibitem[{\citenamefont{Martin et~al.}(2014{\natexlab{d}})\citenamefont{Martin,
  Ringeval, Trotta, and Vennin}}]{Martin:2013nzq}
\bibinfo{author}{\bibfnamefont{J.}~\bibnamefont{Martin}},
  \bibinfo{author}{\bibfnamefont{C.}~\bibnamefont{Ringeval}},
  \bibinfo{author}{\bibfnamefont{R.}~\bibnamefont{Trotta}}, \bibnamefont{and}
  \bibinfo{author}{\bibfnamefont{V.}~\bibnamefont{Vennin}},
  \bibinfo{journal}{JCAP} \textbf{\bibinfo{volume}{1403}}, \bibinfo{pages}{039}
  (\bibinfo{year}{2014}{\natexlab{d}}), \eprint{1312.3529}.

\bibitem[{\citenamefont{Lewis et~al.}(2000)\citenamefont{Lewis, Challinor, and
  Lasenby}}]{Lewis:1999bs}
\bibinfo{author}{\bibfnamefont{A.}~\bibnamefont{Lewis}},
  \bibinfo{author}{\bibfnamefont{A.}~\bibnamefont{Challinor}},
  \bibnamefont{and} \bibinfo{author}{\bibfnamefont{A.}~\bibnamefont{Lasenby}},
  \bibinfo{journal}{Astrophys.J.} \textbf{\bibinfo{volume}{538}},
  \bibinfo{pages}{473} (\bibinfo{year}{2000}), \eprint{astro-ph/9911177}.

\bibitem[{\citenamefont{Lewis and Bridle}(2002)}]{Lewis:2002ah}
\bibinfo{author}{\bibfnamefont{A.}~\bibnamefont{Lewis}} \bibnamefont{and}
  \bibinfo{author}{\bibfnamefont{S.}~\bibnamefont{Bridle}},
  \bibinfo{journal}{Phys.Rev.} \textbf{\bibinfo{volume}{D66}},
  \bibinfo{pages}{103511} (\bibinfo{year}{2002}), \eprint{astro-ph/0205436}.

\bibitem[{\citenamefont{Thacker et~al.}(2010)\citenamefont{Thacker, Zhang,
  Watson, Birch, and Berry}}]{Thacker:2010}
\bibinfo{author}{\bibfnamefont{W.~I.} \bibnamefont{Thacker}},
  \bibinfo{author}{\bibfnamefont{J.}~\bibnamefont{Zhang}},
  \bibinfo{author}{\bibfnamefont{L.~T.} \bibnamefont{Watson}},
  \bibinfo{author}{\bibfnamefont{M.~A.} \bibnamefont{Birch},
  \bibfnamefont{Jeffrey B. an d~Iyer}}, \bibnamefont{and}
  \bibinfo{author}{\bibfnamefont{M.~W.} \bibnamefont{Berry}},
  \bibinfo{journal}{ACM Trans. Math. Softw.} \textbf{\bibinfo{volume}{37}},
  \bibinfo{pages}{34:1} (\bibinfo{year}{2010}), ISSN \bibinfo{issn}{0098-3500},
  \urlprefix\url{http://doi.acm.org/10.1145/1824801.1824812}.

\bibitem[{\citenamefont{{Feroz} and {Hobson}}(2008)}]{Feroz:2007kg}
\bibinfo{author}{\bibfnamefont{F.}~\bibnamefont{{Feroz}}} \bibnamefont{and}
  \bibinfo{author}{\bibfnamefont{M.~P.} \bibnamefont{{Hobson}}},
  \bibinfo{journal}{Mon. Not. R. Astron. Soc.} \textbf{\bibinfo{volume}{384}},
  \bibinfo{pages}{449} (\bibinfo{year}{2008}), \eprint{0704.3704}.

\bibitem[{\citenamefont{{Feroz} et~al.}(2009)\citenamefont{{Feroz}, {Hobson},
  and {Bridges}}}]{Feroz:2008xx}
\bibinfo{author}{\bibfnamefont{F.}~\bibnamefont{{Feroz}}},
  \bibinfo{author}{\bibfnamefont{M.~P.} \bibnamefont{{Hobson}}},
  \bibnamefont{and}
  \bibinfo{author}{\bibfnamefont{M.}~\bibnamefont{{Bridges}}},
  \bibinfo{journal}{Mon. Not. R. Astron. Soc.} \textbf{\bibinfo{volume}{398}},
  \bibinfo{pages}{1601} (\bibinfo{year}{2009}), \eprint{0809.3437}.

\bibitem[{\citenamefont{Feroz et~al.}(2013)\citenamefont{Feroz, Hobson,
  Cameron, and Pettitt}}]{Feroz:2013hea}
\bibinfo{author}{\bibfnamefont{F.}~\bibnamefont{Feroz}},
  \bibinfo{author}{\bibfnamefont{M.}~\bibnamefont{Hobson}},
  \bibinfo{author}{\bibfnamefont{E.}~\bibnamefont{Cameron}}, \bibnamefont{and}
  \bibinfo{author}{\bibfnamefont{A.}~\bibnamefont{Pettitt}}
  (\bibinfo{year}{2013}), \eprint{1306.2144}.

\bibitem[{\citenamefont{Shannon}(1951)}]{Shannon:1951}
\bibinfo{author}{\bibfnamefont{C.~E.} \bibnamefont{Shannon}},
  \bibinfo{journal}{Bell System Technical Journal}
  \textbf{\bibinfo{volume}{30}}, \bibinfo{pages}{50} (\bibinfo{year}{1951}),
  ISSN \bibinfo{issn}{1538-7305}.

\bibitem[{\citenamefont{Scaramella et~al.}(2015)}]{Scaramella:2015rra}
\bibinfo{author}{\bibfnamefont{R.}~\bibnamefont{Scaramella}}
  \bibnamefont{et~al.} (\bibinfo{collaboration}{Euclid}), \bibinfo{journal}{IAU
  Symp.} \textbf{\bibinfo{volume}{306}}, \bibinfo{pages}{375}
  (\bibinfo{year}{2015}), \eprint{1501.04908}.

\bibitem[{\citenamefont{Matsumura et~al.}(2013)}]{Matsumura:2013aja}
\bibinfo{author}{\bibfnamefont{T.}~\bibnamefont{Matsumura}}
  \bibnamefont{et~al.} (\bibinfo{year}{2013}), \bibinfo{note}{[J. Low. Temp.
  Phys.176,733(2014)]}, \eprint{1311.2847}.

\bibitem[{\citenamefont{COrE}()}]{core}
\bibinfo{author}{\bibnamefont{COrE}}, \emph{\bibinfo{title}{{A satellite
  mission for probing cosmic origins, neutrinos masses and the origin of stars
  and magnetic fields}}},
  \urlprefix\url{{http://www.core-mission.org/science.php}}.

\bibitem[{\citenamefont{Garcia-Bellido
  et~al.}(2009)\citenamefont{Garcia-Bellido, Figueroa, and
  Rubio}}]{GarciaBellido:2008ab}
\bibinfo{author}{\bibfnamefont{J.}~\bibnamefont{Garcia-Bellido}},
  \bibinfo{author}{\bibfnamefont{D.~G.} \bibnamefont{Figueroa}},
  \bibnamefont{and} \bibinfo{author}{\bibfnamefont{J.}~\bibnamefont{Rubio}},
  \bibinfo{journal}{Phys. Rev.} \textbf{\bibinfo{volume}{D79}},
  \bibinfo{pages}{063531} (\bibinfo{year}{2009}), \eprint{0812.4624}.

\bibitem[{\citenamefont{Terada et~al.}(2015)\citenamefont{Terada, Watanabe,
  Yamada, and Yokoyama}}]{Terada:2014uia}
\bibinfo{author}{\bibfnamefont{T.}~\bibnamefont{Terada}},
  \bibinfo{author}{\bibfnamefont{Y.}~\bibnamefont{Watanabe}},
  \bibinfo{author}{\bibfnamefont{Y.}~\bibnamefont{Yamada}}, \bibnamefont{and}
  \bibinfo{author}{\bibfnamefont{J.}~\bibnamefont{Yokoyama}},
  \bibinfo{journal}{JHEP} \textbf{\bibinfo{volume}{02}}, \bibinfo{pages}{105}
  (\bibinfo{year}{2015}), \eprint{1411.6746}.

\bibitem[{\citenamefont{Figueroa et~al.}(2015)\citenamefont{Figueroa,
  Garcia-Bellido, and Torrenti}}]{Figueroa:2015rqa}
\bibinfo{author}{\bibfnamefont{D.~G.} \bibnamefont{Figueroa}},
  \bibinfo{author}{\bibfnamefont{J.}~\bibnamefont{Garcia-Bellido}},
  \bibnamefont{and} \bibinfo{author}{\bibfnamefont{F.}~\bibnamefont{Torrenti}},
  \bibinfo{journal}{Phys. Rev.} \textbf{\bibinfo{volume}{D92}},
  \bibinfo{pages}{083511} (\bibinfo{year}{2015}), \eprint{1504.04600}.

\bibitem[{\citenamefont{Vennin et~al.}(2015)\citenamefont{Vennin, Koyama, and
  Wands}}]{Vennin:2015vfa}
\bibinfo{author}{\bibfnamefont{V.}~\bibnamefont{Vennin}},
  \bibinfo{author}{\bibfnamefont{K.}~\bibnamefont{Koyama}}, \bibnamefont{and}
  \bibinfo{author}{\bibfnamefont{D.}~\bibnamefont{Wands}},
  \bibinfo{journal}{JCAP} \textbf{\bibinfo{volume}{1511}}, \bibinfo{pages}{008}
  (\bibinfo{year}{2015}), \eprint{1507.07575}.

\bibitem[{\citenamefont{Vennin et~al.}(2016)\citenamefont{Vennin, Koyama, and
  Wands}}]{Vennin:2015egh}
\bibinfo{author}{\bibfnamefont{V.}~\bibnamefont{Vennin}},
  \bibinfo{author}{\bibfnamefont{K.}~\bibnamefont{Koyama}}, \bibnamefont{and}
  \bibinfo{author}{\bibfnamefont{D.}~\bibnamefont{Wands}},
  \bibinfo{journal}{JCAP} \textbf{\bibinfo{volume}{1603}}, \bibinfo{pages}{024}
  (\bibinfo{year}{2016}), \eprint{1512.03403}.

\end{thebibliography}

\end{document}